\documentclass[aps,10pt,nolongbibliography,prd,tightenlines,twocolumn,twoside,showpacs,superscriptaddress]{revtex4-1}
\usepackage[caption=false]{subfig}

\usepackage[hidelinks]{hyperref}
\usepackage{color}
\usepackage{graphicx}
\usepackage{siunitx}
\usepackage{amsmath,amssymb,amsfonts, amsthm}

\usepackage{bm}
\usepackage{hyperref}
\usepackage{lineno}

\usepackage[normalem]{ulem}  

\renewcommand{\sout}{\bgroup \color{red} \ULdepth=-.5ex \ULset}

\begin{document}
	\title{Spin alignment of $K^\ast$ induced by strange-baryon density inhomogeneity}
	
	\author{Feng Li
	}
        \email{lifengphysics@gmail.com}
	\email{fengli@lzu.edu.cn}
        \affiliation{School of Physics and Electronics, Hunan University, Changsha 410082, China} 
	\affiliation{Hunan Provincial Key Laboratory of High-Energy Scale Physics and Applications, Hunan University, Changsha 410082, China}
	\affiliation{School of Physical Science and Technology, Lanzhou University, Lanzhou 730000, China}
	\affiliation{Lanzhou Center for Theoretical Physics, Key Laboratory of Theoretical Physics of Gansu Province}
	\affiliation{Frontiers Science Center for Rare Isotopes, Lanzhou University}
	
\begin{abstract}
The difference between the spin alignments of $K^\ast$ and those of $\phi$ at the low collision energies is a puzzle raised by the recent experiments. Unlike $\phi$ meson, $K^\ast$, carrying a unit strange charge, should react to strange chemical potential $\mu_S$. In this paper, I shall first convince you that $\mu_S$ is not small in a baryon-rich medium for keeping strange neutrality, and then derive the spin alignment induced by the gradient of $\mu_S$ using linear response theory, with the transport coefficients expressed, without any approximation, in terms of the $K^\ast$'s in-medium spectral properties by employing Ward-Takahashi identity. It turns out that such an effect applies mainly to the particles whose longitudinal and transverse modes diverge, and induces only the local spin alignment in a static medium. The magnitudes of these coefficients will be further estimated under the quasi-particle approximation.
\end{abstract}
	
\maketitle

\section{Introduction}%

The strong magnetic fields and relativistic flows generated in high energy heavy-ion collisions (HICs) induce probably the most abundant spin polarization phenomena on this planet, including chiral magnetic effect~\cite{Kharzeev:2007jp,Fukushima:2008xe}, chiral vortical effect~\cite{Vilenkin:1979ui,Son:2009tf,Kharzeev:2010gr,Gao:2012ix,Jiang:2015cva,Kharzeev:2015znc}, shear induced polarization~\cite{Liu:2021uhn,Fu:2021pok}, and spin Hall effect~\cite{Liu:2020dxg,Fu:2022oup}, which open a new window probing the properties of quark-gluon plasma (QGP). Among these polarization phenomena, the tensor polarization, or spin-alignment, of the vector mesons are under active investigations and discussions. 

Spin alignment, defined as $\delta \rho_{00} \equiv \rho_{00} -1/3$ with $\rho_{00}$ being the probability to find a vector meson in $|0\rangle$, characterizes the tendency for the vector mesons to be in $|\pm 1\rangle$ or $|0\rangle$. Proposed in 2005 based on quark coalescence model, the spin alignment was considered to have originated from the polarization of the constituent quarks~\cite{Liang:2004xn,Liang:2004ph}, and was further estimated with thermal model as $\delta \rho_{00} \sim \mathcal O (\omega^2 / T^2) \sim O (10^{-4})$ in HICs~\cite{Becattini:2007sr} with $\omega$ and $T$ representing the vorticity and temperature of the rotating fireball respectively, under the assumption that the polarization is purely induced due to vortical effect. This estimated value is, however, much smaller than the later on measurements carried out for various vector mesons, including $K^\ast$, $\phi$ and $J/\psi$, in Relativistic Heavy-Ion Collider (RHIC) and Large Hadron Collider (LHC)~\cite{STAR:2008lcm,ALICE:2019aid,Singha:2020qns,STAR:2022fan}.

Substantial efforts have been made to fix the mismatch between the theoretical estimations and experimental measurements, including improving the quark coalescence model~\cite{Sheng:2020ghv}, introducing an exotic fluctuating strong field~\cite{Sheng:2022wsy}, and connecting the spin alignment with the turbulent color fields~\cite{Kumar:2022ylt} or the fluctuating glasma field in the HICs~\cite{Kumar:2023ghs}. Among these theoretical efforts, the discovery of shear induced tensor polarization (SITP) was made recently via both the linear response theory~\cite{Li:2022vmb,Dong:2023cng} and the quantum kinetic theory~\cite{Wagner:2022gza}, and is promising to describe the data obtained in the Au-Au collisions at 200 GeV~\cite{Li:2022vmb}. Furthermore, it is illustrated in Ref.~\cite{Li:2022vmb} that the SITP mechanism bridges two key problems in high energy nuclear physics, i.e., the spin alignment and the in-medium spectral properties of the vector mesons, making the former a potential probe of the latter.

Besides the magnitude, the dependencies of the spin alignments on the transverse momenta ($p_T$), centralities, collisional energies and particle species are puzzling as well. The $p_T$ and centrality dependencies might be promisingly described in the SITP scenario with sincere efforts~\cite{Li:2022vmb}. In this paper, I will focus on the specie dependence of the spin alignments, i.e., the difference between the spin alignments of $K^\ast$ and those of $\phi$, which is enlarged in the low energy collisions. 

Unlike $\phi$ meson, $K^\ast$ carries a unit strange charge, making it reacting to the strange chemical potential $\mu_S$. Therefore, the gradient $\partial(\beta\mu_S)$ could contribute to the spin alignments of $K^\ast$, but not to those of $\phi$. It might make the spin alignments of both the particle species diverge. Since $\partial(\beta\mu_S)$ is usually related to the diffusion of strange charge, I would like to call this effect {\it diffusion induced tensor polarization} (DITP). I would not conclude that the DITP effect is the key machinery leading to the difference between $K^\ast$ and $\phi$'s spin alignments. In fact, it will be pointed out in Section \ref{sec:LocalSA} that {\it the DITP effect only contributes to the \lq\lq local\rq\rq spin alignments in a static medium}, and hence most probably is not the reason for the observed difference on the \lq\lq global\rq\rq spin alignment between the particle species. But still, it is an interesting effect. As explained later, DITP could be affected drastically by the spinodal decomposition~\cite{Li:2015pbv,Li:2016uvu}, which makes it a potential signal of first order phase transition that has been sought for a decade in the beam energy scan (BES) experiments~\cite{BES}. 

Following the avenue paved in Ref.~\cite{Li:2022vmb}, I will show the derivation of DITP using linear response theory, with the corresponding transport coefficients evaluated beyond random phase approximation. By employing Ward-Takahashi identity~\cite{Ward1950}, these coefficients can be expressed non-perturbatively in terms of meson's spectral properties, and further estimated under the quasi-particle approximation. But before the derivation, I would first like to clarify shortly in the next section that $\mu_S$ is NOT negligible in the low energy HICs. 

\section{Strange-Baryon Correspondence}
\label{sec:Strange-Baryon}

The significance of $\mu_S$ in the low energy HICs can be intuitively illustrated by the simulation on the Ar+KCl collisions at 1.76 GeV~\cite{Li:2012bga}, where the produced kaons amount to 50 times the number of anti-kaons, which is reasonable, since in such a baryon rich medium, $s$ quarks are more likely to be bounded with di-quarks into hyperons, rather than paired with anti-quarks into anti-kaons. From the perspective of thermal model, a large ratio $N_K / N_{\bar K} \sim \exp (2\mu_S/T)$ suggests a large $\mu_S$ in these low energy HICs.

\begin{figure}
    \centering
    \includegraphics[width=0.45\textwidth]{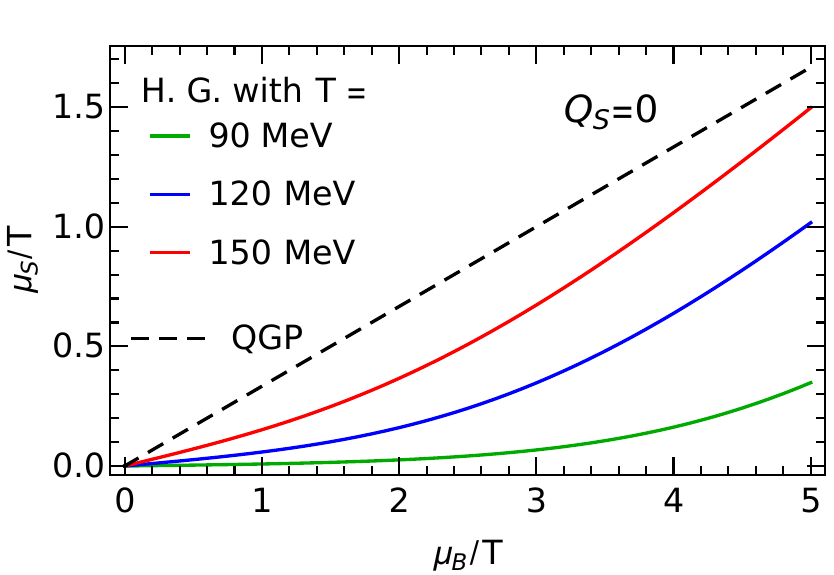}
    \caption{The dependence of strange chemical potential $\mu_S$ on baryon chemical potential $\mu_B$ for a strange neutral system either in partonic phase or in hadronic phase under various temperatures.}
    \label{fig:Strange-Baryon}
\end{figure}

In fact, $\mu_S$ is positively dependent on $\mu_B$ for any strange neutral systems. In partonic phase, given that a $s$ quark carries $1/3$ baryon charge and $-1$ strange charge, $\mu_S$ should be equal to $\mu_B / 3$ so that there are equal numbers of $s$ and $\bar s$ quarks in the system. The contribution from electrical chemical potential is negligible at all collisional energies~\cite{Bhattacharyya:2019cer}, and is hence neglected in the above analysis. For the strange neutral hadron gas composed of all the particle species listed in the PDG notebook~\cite{ParticleDataGroup:2022pth}, the dependencies of $\mu_S$ on $\mu_B$ under various temperatures are plotted by the solid lines in Fig.~\ref{fig:Strange-Baryon}. As shown, in all the thermal equilibrium cases, $\mu_S$ increases with $\mu_B$, in consistent with the results shown in Ref.~\cite{Bhattacharyya:2019cer}, and approaches the partonic limit, i.e., $\mu_B / 3$, under high temperatures. 

The strange-baryon correspondence demonstrated above suggests not only the significance of the driving force of the DITP effect, i.e., $\partial (\beta \mu_S)$, in the low energy HICs, but also the potential connection between the DITP and the first order phase transition, since, as illustrated in Fig.~\ref{fig:Strange-Baryon}, $\partial (\beta \mu_S)$ is positively dependent on $\partial (\beta \mu_B)$, while the latter will be enlarged with density inhomogeneity generated during a first order phase decomposition~\cite{Li:2015pbv,Li:2016uvu}, which is observed and quantified by the density moments in the simulations on the HICs at the BES energies~\cite{Sun:2020pjz,Sun:2022cxp}. It makes the DITP effect a potential novel probe of QCD first order phase transition.

After all these commercials, let us look into the derivation of DITP in detail.

\section{Derivation of DITP}

The derivation of DITP is based on the linear response theory, where $\partial (\beta \mu_S)$ plays the role of the driving force and $\delta \rho_{00}$ is the response of the system proportional to the driving force. Before I go further, let me first express $\delta\rho_{00}$ in terms of the field operators.

\subsection{Observable}

Following Ref.~\cite{Li:2022vmb}, $\delta\rho_{00}$ is again expressed in terms of the Wigner function of the vector mesons. In detail, the {\it single particle} density matrix element $\rho_{ss^\prime}(\mathbf p)$ is defined as $\mathcal N^{-1} a^\dagger_s (\mathbf p) a_{s^\prime}(\mathbf p)$ where $a^\dagger_s$ and $a_s$ are the operators generating and annihilating a vector meson in the state $|s\rangle$, and $\mathcal N$ is the normalization factor keeping $\sum_s \rho_{ss}=1$. The annihilating and generating operators can be further expressed in terms of the field operators (see e.g., Chpt. 5 in Ref.~\cite{Weinberg:1995mt}) using 
\begin{equation}
    \widetilde\phi^\mu (p) = \frac{(2\pi)^{4-\frac{3}{2}}}{\sqrt{2E_p}} \delta(p^0-E_p) \sum_s \epsilon^\mu_s (\mathbf p) a_s(\mathbf p),
\end{equation}
where $E_p = \sqrt{\mathbf p^2 + m^2}$ is the on-shell energy, $\phi$ represents the {\it particle} annihilation field and $\epsilon$ represents the {\it on-shell} covariant polarizer satisfying $\epsilon^\ast_s (\mathbf p)\cdot \epsilon_{s^\prime}(\mathbf p)=-\delta_{ss^\prime}$ and $\epsilon_s (\mathbf p)\cdot \tilde p=0$ with $\tilde p\equiv (E_p, \mathbf p)$ representing the on-shell 4-momentum. It finally leads to
\begin{eqnarray}
    \label{Eq:Wigner}
    \rho_{ss^\prime}(\mathbf p) &=& \mathcal N^{-1} \epsilon^\mu_s(\mathbf p) \int d^3 \mathbf X W_{\mu\nu} (X,\mathbf p) \epsilon^{\nu\ast}_{s^\prime}(\mathbf p)\\
    W_{\mu\nu}(X,\mathbf p) &\equiv& E_p \int \frac{d p^0}{\pi} d^4 y e^{ip\cdot y} \phi^\dagger_\mu\left(X-\frac y 2\right) \phi_\nu \left(X+\frac y 2\right), \nonumber
\end{eqnarray}
where $W$ is the Wigner function characterizing the phase space distribution of the vector mesons, with the integration on $p^0$ covering only the positive and time-like frequencies to exclude the contributions from the anti-particle sectors. The normalization factor can thus be determined, using $\sum_s \epsilon^\mu_s \epsilon^{\nu\ast}_s \equiv \tilde{P}^{\mu\nu}$, as
\begin{equation}
    \mathcal N = \tilde P^{\mu\nu}\int d^3 \mathbf X W_{(\mu\nu)} (X,\mathbf p),
\end{equation}
where $P^{\mu\nu} (p)\equiv - g^{\mu\nu} + p^\mu p^\nu / p^2$ is the projector orthogonal to momentum, satisfying $P\cdot p=0$, $P^\mu_\mu=-3$ and $P\cdot P = -P$, and $\tilde{P}$ is a shorthand of $P(\tilde p)$ satisfying $\tilde P\cdot\epsilon_s=-\epsilon_s$. Given that $\epsilon_s$ is real for $s=0$, the spin alignment can thus be written as
\begin{equation}
    \delta\rho_{00}(\mathbf p) = \mathcal N^{-1} \epsilon^{\langle\mu}_0(\mathbf p) \epsilon^{\nu\rangle}_0(\mathbf p) \int d^3 \mathbf X W_{(\mu\nu)} (X,\mathbf p),
    \label{Eq:SpinAlignmentOperator}
\end{equation}
where the round brackets mean to make the tensor symmetric about the indices in the brackets, i.e., $A_{(\alpha\beta)}\equiv A_{\alpha\beta}/2 + A_{\beta\alpha}/2$, and the angle brackets mean to make the tensor further \lq\lq trace\rq\rq-less, i.e., $A_{\langle \alpha\beta \rangle}\equiv A_{(\alpha\beta)} - \tilde P_{\alpha\beta} \tilde P^{\mu\nu}A_{\mu\nu}/3$, resulting in $\epsilon^{\langle\mu}_0\epsilon^{\nu\rangle}_0 = \epsilon^{\mu}_0\epsilon^{\nu}_0 - \tilde P^{\mu\nu} /3$.

Eq. (\ref{Eq:SpinAlignmentOperator}) exhibits a general dependence of spin alignment on the Wigner function, whose expectation value can be decomposed as $\overline{W} = W^{(0)} + W^{(1)}$ with $W^{(0)}$ and $W^{(1)}$ representing the thermal expectation value of $W$ and its correction proportional to the \lq\lq driving force\rq\rq respectively. After keeping the leading order of $W^{(1)}$, one obtain
\begin{equation}
    \overline{\delta\rho_{00}} = \alpha_0 + \mathcal N_0^{-1} \left[\epsilon^{\langle\mu}_0 \epsilon^{\nu\rangle}_0 - \alpha_0 \tilde P^{\mu\nu} \right]\int d^3 \mathbf X W_{(\mu\nu)}^{(1)},
    \label{Eq:SpinAlignmentExp}
\end{equation}
where
\begin{eqnarray}
    \alpha_0 &=&  \mathcal N_0^{-1} \epsilon^{\langle\mu}_0 \epsilon^{\nu\rangle}_0  \int d^3 \mathbf X W_{(\mu\nu)}^{(0)},\\
    \mathcal N_0 &=& \tilde P^{\mu\nu}\int d^3 \mathbf X W_{(\mu\nu)}^{(0)}\nonumber
\end{eqnarray}
characterizes the {\it zeroth order tensor polarization}, originating from the in-medium divergence between the longitudinal and the transverse modes of the vector mesons, which has been proposed and studied in Ref.~\cite{Li:2022vmb,Dong:2023cng}, and is proven negligible at least for $\phi$ meson~\cite{Dong:2023cng}. In the remaining part of this paper, I shall focus on the second term in Eq. (\ref{Eq:SpinAlignmentExp}) and evaluate $W^{(1)}$ originating from the gradient $\partial(\beta \mu_S)$ using the linear response theory. But before exploring $W^{(1)}$ with linear response theory, let me end this subsection by mentioning the last point, i.e., in a real HIC, the spatial integration $\int d^3 \mathbf X$ in Eq. (\ref{Eq:SpinAlignmentOperator}) and (\ref{Eq:SpinAlignmentExp}) should be replaced with the integration over the hyper-surface at freeze-out. 

\begin{figure}
    \centering
    \includegraphics[width=0.37\textwidth]{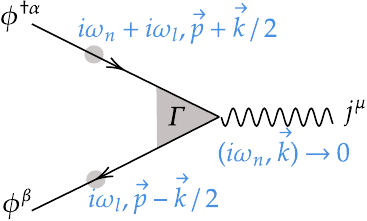}
    \caption{The Feynman diagram for calculating $\varrho^{\alpha\beta;\mu}_{WJ}$ in Eq. (\ref{Eq:RhoWJ}). The fields $\phi^{\dagger\alpha}$ and $\phi^\beta$ are at the same time, resulting in the summation on the Matsubara frequency $i\omega_l$. The momentum flows to $j^\mu$ vanishes.}
    \label{fig:Feynman}
\end{figure}

\subsection{Linear Response}

Following the approach developed by Zubarev~\cite{zubarev1974nonequilibrium,Hosoya:1983id}, $W_{(1)}$ under an in-homogeneous strange charge distribution can be obtained as
\begin{eqnarray}
    W_{(1)}^{(\alpha\beta)} &=& -\lim_{k^0 \to 0} \frac{\varrho_{WJ}^{\alpha\beta;\mu}(k^0,\mathbf k=0)}{k^0} \frac{\partial_\mu(\beta \mu_S)}{\beta}, \\
    \label{Eq:LinearResponse}
    \varrho_{WJ}^{\alpha\beta;\mu} (k) &\equiv& \int d^4 x e^{ik\cdot x} \left[W^{(\alpha\beta)} \left(x^\prime+\frac x 2, \mathbf p\right), j^\mu \left(x^\prime-\frac x 2\right)\right]. \nonumber
\end{eqnarray}

After substituting $W^{\alpha\beta}$ with the field operators using Eq. (\ref{Eq:Wigner}), and transforming the commutators in $\varrho$ into a Matsubara propagator via an analytical continuation (see, e.g., Chpt. 2.4 in Ref.~\cite{Bellac:2011kqa}), $\varrho$ can be further expressed as the three-point Green's function illustrated in Fig. \ref{fig:Feynman} as
\begin{widetext}
\begin{align}
    \label{Eq:RhoWJ}
    \varrho_{WJ}^{\alpha\beta;\mu} (k) &= 4 E_p \int d^3\mathbf y e^{-i \mathbf p\cdot \mathbf y} \int_0^\beta d^4 x e^{i\omega_n x^0-i\mathbf k\cdot \mathbf x}\left. \mathrm{Im} \left\langle \mathcal T_\tau \phi^{\dagger (\alpha} \left(x^\prime+\frac x 2 - \frac {\mathbf y} 2\right)\phi^{\beta)} \left(x^\prime+\frac x 2 + \frac {\mathbf y} 2\right)j^\mu\left(x^\prime-\frac x 2\right)\right\rangle\right|_{i\omega_n \to k^0 + i0^+}\nonumber\\
    &=4E_p T \sum_l \mathrm{Im} \left[\widetilde S^{(\beta\rho}\left(i\omega_n+i\omega_l, \mathbf p +\frac{\mathbf k}{2}\right) \widetilde \Gamma^\mu_{\rho\sigma}\left(i\omega_n+i\omega_l,\mathbf p+\frac{\mathbf k}{2}; i\omega_l, \mathbf p -\frac{\mathbf k}{2}\right)\widetilde S^{\sigma\alpha)}\left(i\omega_l, \mathbf p -\frac{\mathbf k}{2}\right)\right]\\
    &=2E_p \int_{C_\uparrow+C_\downarrow} \frac{d p^0}{2\pi i} \mathrm{Im} \left[\widetilde S^{(\beta\rho}\left(i\omega_n+p^0, \mathbf p +\frac{\mathbf k}{2}\right) \widetilde \Gamma^\mu_{\rho\sigma}\left(i\omega_n+p^0,\mathbf p+\frac{\mathbf k}{2}; p^0, \mathbf p -\frac{\mathbf k}{2}\right)\widetilde S^{\sigma\alpha)}\left(p^0, \mathbf p -\frac{\mathbf k}{2}\right)\right]\coth \frac{p^0}{2T}, \nonumber
\end{align}
\end{widetext}
where $\widetilde S$ and $\widetilde \Gamma$ represent the {\it non-perturbative} meson propagator and vertex for conserved current respectively, with the Matsubara frequencies $\omega_n$ and $\omega_l$ taking the discrete values, i.e., $2\pi n T$ and $2\pi l T$. The transition from the summation over $\omega_l$ to the integration over $p^0$ at the last equality, with $C_\uparrow$ and $C_\downarrow$ representing the upward and downward integration contours slightly left and right to the imaginary axis respectively, is based on the Residue's theorem as explained in Ref.~\cite{Bellac:2011kqa}.

Eq. (\ref{Eq:RhoWJ}) is EXACT, without any approximation. The non-perturbative propagator $\widetilde S$ can be further written in the spectral representation (see, e.g., Chpt. 2.4 of Ref.~\cite{Bellac:2011kqa}) as
\begin{equation}
    \label{Eq:SpectralRepresentation}
    \widetilde S^{\alpha\beta} (i\omega_n, \mathbf p) = \int \frac{dp^0}{2\pi} \sum_{a=L,T} \frac{P^{\alpha\beta}_a (p) \mathcal A_a (p)}{i\omega_n-p^0}, 
\end{equation}
where $\mathcal A_L$ and $\mathcal A_T$ are the spectral functions of the longitudinal and transverse modes of the vector mesons, and $P_L$ and $P_T$ defined in Eq. (\ref{Eq:LTProjector}) are the longitudinal and transverse projectors respectively. In vacuum, one should expect $\mathcal A_L = \mathcal A_T \propto \delta (p^0 - E_p)$ for stable particles. However, in a thermal medium, the spectral functions shall be broadened with their peak locations being shifted. Furthermore, $\mathcal A_L$ and $\mathcal A_T$ might diverge as well, with the difference vanishing for the zero-momentum particles in the medium rest frame (MRF). As demonstrated in Appendix \ref{sec:AppdxProjector}, \lq\lq longitudinal\rq\rq and \lq\lq transverse\rq\rq mean being parallel and perpendicular to the particle's 3-momentum respectively in the MRF, so both the modes should degenerate for the particles with vanishing 3-momenta.

Under the assumption that {\it there is no pole of $p^0$ in the vertex $\widetilde \Gamma^\mu_{\rho\sigma}(i\omega_n+p^0,\mathbf p+\frac{\mathbf k}{2}; p^0, \mathbf p -\frac{\mathbf k}{2})$}, the integration over $p^0$ in Eq. (\ref{Eq:RhoWJ}) can be conducted by collecting only the residues of  the poles in $\widetilde S$ indicated in the denominator of Eq. (\ref{Eq:SpectralRepresentation}), which results in a simple relation as follows,
\begin{widetext}
    \begin{align}
        \label{Eq:W1}
        W_{(1)}^{(\alpha\beta)} = & \left(\Theta^{\alpha\beta;\mu}+\Upsilon^{\alpha\beta;\mu}\right) \frac{\partial_\mu(\beta \mu_S)}{T}, \nonumber\\
        \Theta^{\alpha\beta;\mu}  = & T^2 E_p \int \frac{dp^0}{\pi} \sum_{a,b =L}^T \mathcal A_a (p) \mathcal A_b (p) P_a^{(\alpha\sigma}(p) P_b^{\beta)\rho} (p) \left[ {\rm Re} \widetilde \Gamma^\mu_{\rho\sigma}(p;p)\right] \frac{\partial n}{\partial p^0},\\
        \Upsilon^{\alpha\beta;\mu} = & T^2 E_p \mathcal P \int \frac{d\bar\omega d\bar\omega^\prime}{\pi^2} \sum_{a,b=L}^T P_b^{(\alpha\sigma}(\bar \omega^\prime, \mathbf p) P_a^{\beta)\rho}(\bar \omega, \mathbf p) \frac{\mathcal A_a (\bar \omega, \mathbf p) \mathcal A_b (\bar \omega^\prime, \mathbf p)}{\bar \omega - \bar \omega^\prime} \nonumber\\
        & \times\lim_{k^0\to 0} \frac{1}{2k^0} {\rm Im} \left[\widetilde \Gamma^\mu_{\rho\sigma}(\bar \omega, \mathbf p; \bar \omega -k^0-i0^+, \mathbf p)\coth \frac{\bar\omega}{2T} - \widetilde \Gamma^\mu_{\rho\sigma}( \bar \omega^\prime + k^0 + i0^+, \mathbf p;\bar \omega^\prime, \mathbf p)\coth \frac{\bar\omega^\prime}{2T}\right],\nonumber
    \end{align}
where $n$ is the Bose distribution function and $\mathcal P$ represents evaluating the principle value of the integration. Notice that both $\Theta$ and $\Upsilon$ are dimensionless.

The contribution from $\Upsilon^{\alpha\beta;\mu}$ in Eq. (\ref{Eq:W1}) vanishes if {\it the vertex $\widetilde \Gamma$ is TIME REVERSAL SYMMETRIC and $\partial_{q^0} \widetilde\Gamma^\mu_{\sigma\rho}(p+q; p) |_{q\to 0}$ is REAL} or if {\it $\mathcal A (\omega, \mathbf p)$ is ANALYTIC in the closed upper half-plane of $\omega$ and REAL on the real axis}, where the latter is similar to the assumption employed in the QCD sum rules~\cite{Colangelo:2000dp}. To see the validity of the above statement, one can exchange $\bar\omega$ and $\bar\omega^\prime$ in the second term of $\Upsilon^{\alpha\beta;\mu}$, and obtains
\begin{align}
    \label{Eq:UpsilonTerm}
    \Upsilon^{\alpha\beta;\mu} = & T^2 E_p \sum_{a,b=L}^T \mathcal P \int \frac{d\bar\omega }{\pi}   P_a^{(\alpha\rho}(\bar \omega, \mathbf p) \mathcal A_a (\bar \omega, \mathbf p) \mathcal P \int \frac{ d\bar\omega^\prime}{\pi} P_b^{\beta)\sigma}(\bar \omega^\prime, \mathbf p) \frac{ \mathcal A_b (\bar \omega^\prime, \mathbf p)}{\bar \omega - \bar \omega^\prime}\nonumber\\
    & \times \lim_{k^0\to 0} \frac{1}{2k^0} {\rm Im} \left[\widetilde \Gamma^\mu_{\rho\sigma}(\bar \omega, \mathbf p; \bar \omega -k^0-i0^+, \mathbf p) + \widetilde \Gamma^\mu_{\sigma\rho}( \bar \omega + k^0 + i0^+, \mathbf p;\bar \omega, \mathbf p)\right] \coth \frac{\bar\omega}{2T}\nonumber\\
    = & -T^2 E_p \sum_{a,b=L}^T \mathcal P \int \frac{d\bar\omega }{\pi}   P_a^{(\alpha\rho}(\bar \omega, \mathbf p) \mathcal A_a (\bar \omega, \mathbf p) \mathcal P \int \frac{ d\bar\omega^\prime}{\pi} P_b^{\beta)\sigma}(\bar \omega^\prime, \mathbf p) \frac{ \mathcal A_b (\bar \omega^\prime, \mathbf p)}{\bar \omega - \bar \omega^\prime}\nonumber\\
    & \times \partial_{k^0} {\rm Im}\widetilde\Gamma^\mu_{\sigma\rho}(\bar\omega+k^0, \mathbf p; \bar\omega, \mathbf p)\Big|_{k^0\to 0} \coth \frac{\bar\omega}{2T}.
\end{align}
\end{widetext}
The last equality in Eq. (\ref{Eq:UpsilonTerm}) holds since $\widetilde \Gamma^\mu_{\rho\sigma}(\bar \omega, \mathbf p; \bar \omega -k^0-i0^+, \mathbf p) = \widetilde \Gamma^\mu_{\sigma\rho}( \bar \omega -k^0-i0^+, \mathbf p ; \bar \omega, \mathbf p)$ due to {\it time reversal symmetry}, and is further approximately equal to $\widetilde \Gamma^\mu_{\sigma\rho}(\bar \omega + k^0-i0^+, \mathbf p; \bar \omega, \mathbf p) -2k^0 \partial_{k^0} \widetilde\Gamma^\mu_{\sigma\rho}(p+k; p)|_{k\to 0}$ where $\widetilde \Gamma^\mu_{\sigma\rho}(\bar \omega + k^0-i0^+, \mathbf p; \bar \omega, \mathbf p)$ happens to be the complex conjugate of $\widetilde \Gamma^\mu_{\sigma\rho}( \bar \omega + k^0 + i0^+, \mathbf p;\bar \omega, \mathbf p)$. Eq. (\ref{Eq:UpsilonTerm}) hence vanishes if $\partial_{q^0} \widetilde\Gamma^\mu_{\sigma\rho}(p+q; p)|_{q\to 0}$ is real, which holds at least in vacuum with $p$ and $p+q$ being both on-shell~\cite{Bhagwat:2006pu,Xu:2019ilh}.

On the other hand, the integration over $\bar\omega^\prime$ in Eq. (\ref{Eq:UpsilonTerm}) is condensed in the following short expression
\begin{equation*}
    \mathcal P \int \frac{ d\bar\omega^\prime}{\pi} P_b^{\beta\sigma}(\bar \omega^\prime, \mathbf p) \frac{ \mathcal A_b (\bar \omega^\prime, \mathbf p)}{\bar \omega - \bar \omega^\prime},
\end{equation*}
which, according to Kramers-Kronig relation or dispersion relation (see, e.g., Chpt. 10.8 in Ref.~\cite{Weinberg:1995mt}), can be evaluated as 
\begin{align*}
    \mathcal P \int \frac{ d\bar\omega^\prime}{\pi} P_b^{\beta\sigma}(\bar \omega^\prime, \mathbf p) \frac{ {\rm Re}\mathcal A_b (\bar \omega^\prime, \mathbf p)}{\bar \omega - \bar \omega^\prime} =& - P_b^{\beta\sigma}(\bar \omega, \mathbf p) {\rm Im}\mathcal A_b (\bar \omega, \mathbf p), \\
    \mathcal P \int \frac{ d\bar\omega^\prime}{\pi} P_b^{\beta\sigma}(\bar \omega^\prime, \mathbf p) \frac{ {\rm Im}\mathcal A_b (\bar \omega^\prime, \mathbf p)}{\bar \omega - \bar \omega^\prime} =&  P_b^{\beta\sigma}(\bar \omega, \mathbf p) {\rm Re}\mathcal A_b (\bar \omega, \mathbf p),
\end{align*}
if $\mathcal A (\omega, \mathbf p)$ is analytic in the closed upper half-plane of $\omega$, and vanishes, leading to a vanishing $\Upsilon^{\alpha\beta;\mu}$ as well, if I further require ${\rm Im}\mathcal A (\bar\omega, \mathbf p) =0$, which sounds reasonable. 

So, in the remaining part of the paper, I will get rid of the contribution from $\Upsilon$, and focus on the contribution
\begin{equation}
    \label{Eq:W1Simple}
    W_{(1)}^{(\alpha\beta)} = \Theta^{\alpha\beta;\mu} \frac{\partial_\mu(\beta \mu_S)}{T}.
\end{equation}

\subsection{Non-perturbative Vertex}

To further evaluate $\Theta^{\alpha\beta;\mu}$ in Eq. (\ref{Eq:W1}), one need to know the detailed expression of the vertex $\widetilde \Gamma$ dressed by the thermal or quantum loops. Fortunately enough, one do not need to list and calculate the Feynman diagrams for the dressed vertex order by order, rather, the vertex  $\widetilde \Gamma$ can be determined using Ward-Takahashi identity~\cite{Ward1950,Takahashi1957}. 

Based on the conservation law and the canonical commutation relation, the vertex of $j^\mu$ is linked with the propagator of the charged particle $\widetilde S$ as
\begin{equation}
    \label{Eq:WardIdentity}
    \widetilde \Gamma^\mu_{\rho\sigma} (p;p) = \mathfrak q \partial_{p_\mu} \widetilde S^{-1}_{\rho\sigma}(p),
\end{equation}
with $\mathfrak q$ representing the particle charge. For vector mesons, 
\begin{equation}
    \label{Eq:PropagatorNP}
    \widetilde S^{-1}_{\mu\nu}(p) = -p^2 P_{\mu\nu}-m^2 g_{\mu\nu}-\Pi_L P^L_{\mu\nu} - \Pi_T P^T_{\mu\nu},
\end{equation}
where $\Pi_L$ and $\Pi_T$ represent the longitudinal and transverse self-energies respectively, which are more fundamental than the spectral functions. In fact, both $\mathcal A_L$ and $\mathcal A_T$ can be expressed in terms of the self-energies as
\begin{equation}
    \label{Eq:SpectralExpression}
    \mathcal A_{L/T} (p) = \frac{2{\rm Im} \Pi_{L/T}}{(p^2-m^2+{\rm Re}\Pi_{L/T})^2 + {\rm Im}\Pi_{L/T}^2}.
\end{equation}
Since both $\Pi_L$ and $\Pi_T$ are Lorentz scalars, they should be expressed as the functions of two independent scalars, i.e., $\Pi_{L/T} = \Pi_{L/T}(\varepsilon,\varsigma)$ with $\varepsilon \equiv u\cdot p$ and $\varsigma \equiv p^2$, for $u$ representing the flow velocity. In the MRF, $\varepsilon = p^0$ and $\kappa^2\equiv\varepsilon^2-\varsigma = \mathbf p^2$ represent the energy and square of the 3-momentum, respectively. Hence, the momentum gradient, presenting in Eq. (\ref{Eq:WardIdentity}), on the self-energies are
\begin{equation}
    \label{Eq:DPi}
    \partial_{p_\mu} \Pi_{L/T} = \partial_\varepsilon \Pi_{L/T} u^\mu + 2\partial_\varsigma\Pi_{L/T} p^\mu.
\end{equation}

\subsection{General Form of DITP}

After substituting $\widetilde \Gamma^\mu_{\rho\sigma}$ in Eq. (\ref{Eq:W1}) with Eq. (\ref{Eq:WardIdentity}), (\ref{Eq:PropagatorNP}) and (\ref{Eq:DPi}), and employing the last identity in Eq. (\ref{Eq:ProjectorIdentities}) and then Eq. (\ref{Eq:DPT}), one obtains
\begin{widetext}
\begin{align}
    \label{Eq:ThetaFactor}
    \Theta^{\alpha\beta;\mu} =& -\mathfrak q T^2 E_p \int \frac{dp^0}{\pi} \sum_{a,b =L}^T \mathcal A_a  \mathcal A_b   \left\{ \left[2 p^\mu (1 +\partial_\varsigma {\rm Re} \Pi_a) + u^\mu \partial_\varepsilon {\rm Re}\Pi_a\right]P_a^{\alpha\beta} \delta_{ab}  +{\rm Re}(\Pi_T-\Pi_L) P_a^{(\alpha\sigma} P_b^{\beta)\rho} \partial_{p_\mu}  P^T_{\rho\sigma}\right\} \frac{\partial n}{\partial p^0} \nonumber\\
    =& -\mathfrak q T^2 E_p \int \frac{dp^0}{\pi}  \left\{ 
    \begin{tabular}{c}
        $\mathcal A_L^2 \left[2 p^\mu (1 +\partial_\varsigma {\rm Re}\Pi_L) + u^\mu \partial_\varepsilon {\rm Re} \Pi_L\right]P_L^{\alpha\beta}$ \\
        \(+ \mathcal A_T^2 \left[2 p^\mu (1 +\partial_\varsigma {\rm Re}\Pi_T) + u^\mu \partial_\varepsilon {\rm Re} \Pi_T\right]P_T^{\alpha\beta}\)   \\
        $+2 \varepsilon \kappa^{-2} \mathcal A_L \mathcal A_T {\rm Re} (\Pi_T-\Pi_L) u_\perp^{(\beta} \tilde P^{\alpha) \mu}_T$
    \end{tabular}
    \right\} \frac{\partial n}{\partial p^0}.
\end{align} 

Let me further liberate all the tensors with indices in Eq. (\ref{Eq:ThetaFactor}) out of the integration by transforming the off-shell tensors and vectors, including $p$, $u_\perp$ and $P_{L/T}$, into the on-shell ones using Eq. (\ref{Eq:PTOffShell}), (\ref{Eq:PLOffShell}) and $(\ref{Eq:UperpOffShell})$. Given that the $\alpha$, $\beta$ indices in Eq. (\ref{Eq:ThetaFactor}) will eventually be contracted with those of the {\it on-shell} polarizers and projectors as demonstrated in Eq. (\ref{Eq:SpinAlignmentExp}), $\Theta^{\alpha\beta;\mu}$ can be {\it equivalently} decomposed as
\begin{equation}
    \label{Eq:ThetaDecomposition}
    \Theta^{\alpha\beta;\mu} = \left(\vartheta_L^p \frac{\tilde p^\mu}{m} +\vartheta_L^u u^\mu\right) \tilde P^{\alpha\beta}_L + \left(\vartheta_T^p \frac{\tilde p^\mu}{m} +\vartheta_T^u u^\mu\right) \tilde P^{\alpha\beta}_T + \vartheta_\Delta u^{(\alpha} \tilde P_T^{\beta)\mu},
\end{equation}
where the {\it dimensionless scalar} transport coefficients, whose values can be evaluated in the MRF, are
\begin{align}
    \label{Eq:CoefficientsEXACT}
    \vartheta_L^p = & -2 \mathfrak q m T^2 E_p \int \frac{d\varepsilon}{\pi} \mathcal A_L^2  (1 +\partial_\varsigma {\rm Re}\Pi_L) \frac{\partial n}{\partial \varepsilon} \left(1+ \frac{\Omega^2 \kappa^2}{m^2 \varsigma}\right),\nonumber\\
    \vartheta_T^p = & -2 \mathfrak q m T^2 E_p \int \frac{d\varepsilon}{\pi} \mathcal A_T^2  (1 +\partial_\varsigma {\rm Re}\Pi_T) \frac{\partial n}{\partial \varepsilon}, \nonumber\\
    \vartheta_L^u = & -\mathfrak q  T^2 E_p \int \frac{d\varepsilon}{\pi} \mathcal A_L^2  (2\Omega +2\Omega\partial_\varsigma {\rm Re}\Pi_L + \partial_\varepsilon {\rm Re}\Pi_L) \frac{\partial n}{\partial \varepsilon} \left(1+ \frac{\Omega^2 \kappa^2}{m^2 \varsigma}\right),\\
    \vartheta_T^u = & -\mathfrak q T^2 E_p \int \frac{d\varepsilon}{\pi} \mathcal A_T^2  (2\Omega +2\Omega\partial_\varsigma {\rm Re}\Pi_T + \partial_\varepsilon {\rm Re}\Pi_T) \frac{\partial n}{\partial \varepsilon},\nonumber\\
    \vartheta_\Delta = & 2\mathfrak q T^2 E_p \int \frac{d\varepsilon}{\pi} \frac{\varepsilon}{\kappa^2} \mathcal A_L \mathcal A_T {\rm Re} (\Pi_T-\Pi_L)\frac{\partial n}{\partial \varepsilon} \left(1-\frac{\varepsilon \Omega}{\varsigma}\right),\nonumber
\end{align}
with $\Omega\equiv u\cdot (p-\tilde p)$ characterizing the deviation of the particle's energy $p^0$ from its on-shell kinetic energy $E_p$.

After substituting $W^{(1)}$ in Eq. (\ref{Eq:SpinAlignmentExp}) with Eq. (\ref{Eq:W1Simple}) and (\ref{Eq:ThetaDecomposition}), and employing the identities in Eq. (\ref{Eq:ProjectorIdentities}) and (\ref{Eq:PTPLOnshell}), one should finally obtain the spin alignment as
\begin{align}
    \label{Eq:SpinAlignmentDITP}
    \overline{\delta\rho_{00}} = & \alpha_0 + \mathcal N_0^{-1}   \int d^3 \mathbf X  \left[\left(\vartheta_{sp.}^p \frac{\tilde p^\mu}{m} +\vartheta_{sp.}^u u^\mu\right) \epsilon_{\langle\alpha}^0 \epsilon_{\beta\rangle}^0\tilde P^{\alpha\beta}_L  + \vartheta_\Delta \epsilon_{\alpha}^0 \epsilon_{\beta}^0u^{(\alpha} \tilde P_T^{\beta)\mu}-\alpha_0\left(\vartheta_{tt.}^p \frac{\tilde p^\mu}{m} +\vartheta_{tt.}^u u^\mu\right)\right]  \frac{\partial_\mu(\beta \mu_S)}{T},\\
    \vartheta^p_{sp.} \equiv & \vartheta^p_L - \vartheta^p_T,~~~~\vartheta^p_{tt.} \equiv  \vartheta^p_L + 2 \vartheta^p_T,~~~~ \vartheta^u_{sp.} \equiv  \vartheta^u_L - \vartheta^u_T,~~~~\vartheta^u_{tt.} \equiv  \vartheta^u_L + 2 \vartheta^u_T. \nonumber
\end{align}
\end{widetext}
Again, the spatial integration $\int d^3 \mathbf X$ should be replaced with the integration over the freeze-out hyper-surface. Notice that all the non-vanishing transport coefficients in Eq. (\ref{Eq:SpinAlignmentDITP}), including $\alpha_0$, $\vartheta_{sp.}^{p/u}$ and $\vartheta_\Delta$, are mainly originating from the splitting of the spectral properties between the longitudinal and the transverse modes. Hence, {\it the DITP effect applies mainly to the vector mesons whose longitudinal and transverse modes are different}.

Eq. (\ref{Eq:SpinAlignmentDITP}) and Eq. (\ref{Eq:CoefficientsEXACT}) demonstrate the key  discovery of this work. In the next two sections, I shall look into the tensor structures in Eq. (\ref{Eq:SpinAlignmentDITP}) in detail in the MRF, and further estimate the transport coefficients in Eq. (\ref{Eq:CoefficientsEXACT}) under quasi-particle approximation.

\section{Contribution to the Global and Local Spin Alignment}
\label{sec:LocalSA}

\begin{figure}
    \centering
    \includegraphics[width=0.48\textwidth]{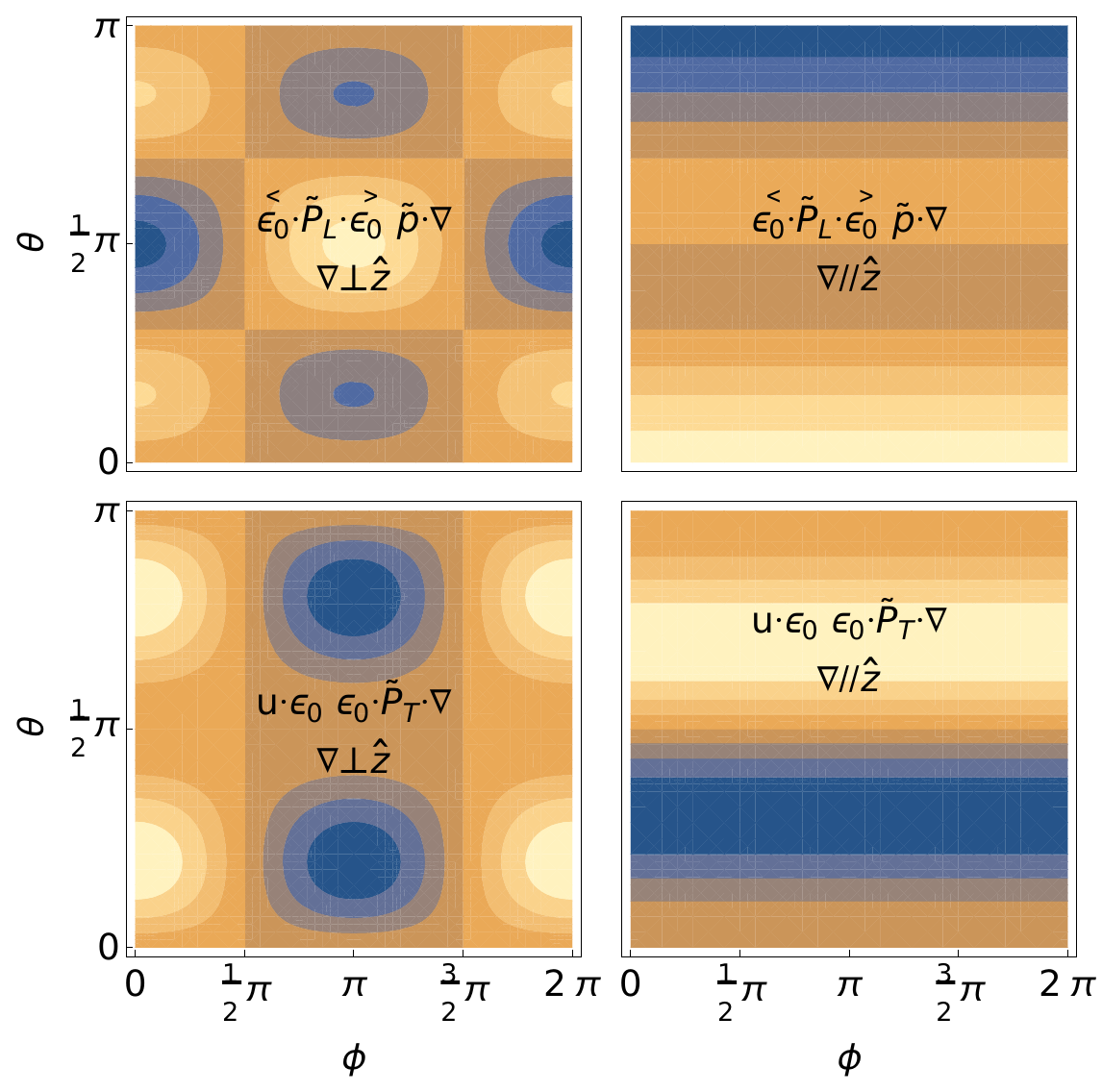}
    \caption{The momentum orientation dependence of the spin alignment induced by the DITP effect contributed by the tensor structure $\epsilon_{\langle\alpha}^0 \epsilon_{\beta\rangle}^0\tilde P^{\alpha\beta}_L \tilde p \cdot \partial$ (upper panels) and $\epsilon_{\alpha}^0 \epsilon_{\beta}^0 u^{(\alpha} \tilde P_T^{\beta)\mu}\partial_\mu$ (lower panels) in a static and near-equilibrium medium, with the chemical potential gradient $\nabla(\beta \mu_S)$ being perpendicular (left column) and parallel (right column) to the spin quantization axis $\hat{\mathbf z}$, respectively, where $\theta$ represents the angle between $\mathbf p$ and $\hat{\mathbf z}$, and $\phi$ represents the azimuth relative to $\hat{\mathbf z}$. A more general orientation dependence of the spin alignment induced by DITP is the superposition of these patterns.}
    \label{fig:OrientationDependence}
\end{figure}

I now evaluate the contributions from each term in Eq. (\ref{Eq:SpinAlignmentDITP}) to the \lq\lq global\rq\rq spin alignment, i.e., the spin alignment obtained after integrating out the particles' 3-momentum, or at least integrating out the orientation of the 3-momentum.  One should expect that some terms may vanish after the momentum (or orientation) integration due to symmetry. Let us figure out who they are.

In the MRF, under the assumption that all the transport coefficients are functions of $\mathbf p^2$ near thermal equilibrium, there are three tensor structures in Eq. (\ref{Eq:SpinAlignmentDITP}) relevant to the momentum orientation, i.e., $\epsilon_{\langle\alpha}^0 \epsilon_{\beta\rangle}^0\tilde P^{\alpha\beta}_L \tilde p \cdot \partial(\beta \mu_S)$, $\epsilon_{\langle\alpha}^0 \epsilon_{\beta\rangle}^0\tilde P^{\alpha\beta}_L u \cdot \partial(\beta \mu_S)$ and $\epsilon_{\alpha}^0 \epsilon_{\beta}^0 u^{(\alpha} \tilde P_T^{\beta)\mu}\partial_\mu(\beta\mu_S)$. The terms proportional to $\alpha_0$ do not contribute additional tensor structures, since $\alpha_0 \propto \epsilon_{\langle\alpha}^0 \epsilon_{\beta\rangle}^0\tilde P^{\alpha\beta}_L$~\cite{Li:2022vmb}.

The polarizer  $\epsilon_0$ can be expressed explicitly in the MRF as
\begin{equation}
    \epsilon_0 = \frac{1}{m} \left(
    \begin{tabular}{c c}
        $E_p$   & $\mathbf p$ \\
        $\mathbf p$  & $m \mathbb I + (E_p-m) \hat{\mathbf p} \hat{\mathbf p}$
    \end{tabular}
    \right) \left(
    \begin{tabular}{c}
         $0$  \\
         $\hat{\mathbf z}$
    \end{tabular}
    \right),
\end{equation}
where $\hat{\mathbf p} \equiv \mathbf p / |\mathbf p|$, and $\hat{\mathbf z}$ is the spin quantization axis in the particle rest frame (PRF). Hence, in the MRF,
\begin{align}
    \label{Eq:TensorStructure}
    \epsilon_{\langle\alpha}^0 \epsilon_{\beta\rangle}^0\tilde P^{\alpha\beta}_L \tilde p \cdot \partial = & \left[\frac{(\mathbf p\cdot \hat{\mathbf z})^2}{\mathbf p^2} -\frac 1 3\right] \left(E_p \partial_t + \mathbf p\cdot \nabla\right),
    \nonumber\\
    \epsilon_{\langle\alpha}^0 \epsilon_{\beta\rangle}^0\tilde P^{\alpha\beta}_L u \cdot \partial = & \left[\frac{(\mathbf p\cdot \hat{\mathbf z})^2}{\mathbf p^2} - \frac 1 3\right] \partial_t, \\
    \epsilon_{\alpha}^0 \epsilon_{\beta}^0 u^{(\alpha} \tilde P_T^{\beta)\mu}\partial_\mu =& -\frac{(\mathbf p \cdot \hat{\mathbf z})}{m} [\hat{\mathbf p}\times(\hat{\mathbf z}\times \hat{\mathbf p})]\cdot \nabla.\nonumber
\end{align}
All the above terms vanish after the momentum or orientation integration, indicating that {\it the DITP effect induces only the \lq\lq local\rq\rq spin alignment in a static medium}, which is illustrated intuitively in Fig. \ref{fig:OrientationDependence}, where the orientation dependencies of the spin alignments in a static medium contributed by the tensor structures listed in Eq. (\ref{Eq:TensorStructure}) are plotted for the cases with $\nabla(\beta\mu_S)$ being perpendicular (left column) and parallel (right column) to $\hat{\mathbf z}$ respectively. A more general orientation dependencies of the spin alignment induced by DITP in a static medium is the superposition of the patterns illustrated in Fig. \ref{fig:OrientationDependence}. These orientation dependencies, apparently different from those generated in the SITP~\cite{Li:2022vmb} effect, may be regarded as the signal of the DITP effect and, as demonstrated in section \ref{sec:Strange-Baryon}, a potential probe of the first order phase transition. 

It should be noted that the above discussions apply only to cases where the spin alignments are counted in a synchronous bulk. If they are counted on a realistic freeze-out hyper-surface, which means the spatial integration in Eq. (\ref{Eq:SpinAlignmentDITP}) is replaced by $\int d\Sigma_\alpha p^\alpha$~\cite{Liu:2020dxg,Li:2022vmb}, the DITP effect might contribute to the \lq\lq global\rq\rq spin alignment due to the extra orientation dependence in the integration measure.

\section{DITP under quasi-particle approximation}

Although the transport coefficients has already been given in Eq. (\ref{Eq:CoefficientsEXACT}), and can thus be evaluated by employing the self-energies and spectral functions of the vector mesons obtained systematically via a non-perturbative formalism such as functional renormalization group (FRG) method~\cite{Jung:2016yxl}, it is still meaningful to express these coefficients in terms of the quantities with intuitive physical meanings, such as spectral width and mass-shift, under the quasi-particle approximation.

Under the quasi-particle approximation, the self-energies are modeled as 
\begin{align}
    {\rm Im} \Pi_{L/T} &\sim E_p \Gamma_{L/T},\nonumber\\
    {\rm Re} \Pi_{L/T} &\sim -2E_p \mathcal E_{L/T}+2\mu_S\varepsilon+\mu_S^2,
\end{align}
with $\Gamma$ and $\mathcal E$, depending only on $\mathbf p^2$ in the MRF, representing the width and mass-shift respectively. The terms with $\mu_S$ originate from transforming $p^2 \to (p^0+\mu)^2-\mathbf p^2$ in the denominator of Eq. (\ref{Eq:SpectralExpression}) in the presence of finite chemical potential. The momentum derivatives presenting in Eq. (\ref{Eq:DPi}) reduce to
\begin{align*}
    1 +\partial_\varsigma {\rm Re}\Pi & = 1 - \mathcal F,\\
    2\Omega +2\Omega\partial_\varsigma {\rm Re}\Pi + \partial_\varepsilon {\rm Re}\Pi & = 2 (\Omega + \mu_S + E_p \mathcal F),
\end{align*}
accordingly, with $\mathcal F \equiv 2 \partial_{\mathbf p^2} (E_p \mathcal E)$. The spectral function can thus be written approximately in the Breit-Wigner form as
\begin{equation}
    \label{Eq:Breit-Wigner}
    \mathcal A_{L/T} \approx \frac{2E_p \Gamma_{L/T}}{4E_p^2 (\Omega+\mu_S-\mathcal E_{L/T})^2 + E_p^2 \Gamma^2_{L/T}},
\end{equation}
which satisfies
\begin{align}
    \label{Eq:Moments}
    E_p \int \frac{d\varepsilon}{\pi} \mathcal A^2_{L/T} =& \frac{1}{E_p \Gamma_{L/T}},\nonumber\\
    E_p \int \frac{d\varepsilon}{\pi} \Omega \mathcal A^2_{L/T} =& \frac{\mathcal E^\prime_{L/T}}{E_p \Gamma_{L/T}},\nonumber\\
    E_p \int \frac{d\varepsilon}{\pi} \Omega^2 \mathcal A^2_{L/T} =& \frac{\frac{1}{4} \Gamma^2_{L/T}+\mathcal E^{\prime 2}_{L/T}}{E_p \Gamma_{L/T}},\\
    E_p \int \frac{d\varepsilon}{\pi} \mathcal A_L \mathcal A_T =& \frac{\bar \Gamma}{E_p \left(\bar\Gamma^2+\mathcal E_\Delta^2\right)},\nonumber\\
    E_p \int \frac{d\varepsilon}{\pi} \Omega \mathcal A_L A_T =& \frac{\bar \Gamma \Bar{\mathcal E}^\prime-\frac 1 4 \Gamma_\Delta \mathcal E_\Delta}{E_p \left(\bar\Gamma^2+\mathcal E_\Delta^2\right)},\nonumber\\
    E_p \int \frac{d\varepsilon}{\pi} \Omega^2 \mathcal A_L A_T =& \frac{\Gamma_L \Gamma_T \bar \Gamma + 2 \mathcal E_T^{\prime 2} \Gamma_L + 2 \mathcal E_L^{\prime 2} \Gamma_T}{4 E_p \left(\bar\Gamma^2+\mathcal E_\Delta^2\right)},\nonumber
\end{align}
where $\bar \Gamma \equiv (\Gamma_L+\Gamma_T) / 2$ and $\Bar{\mathcal E} \equiv (\mathcal E_L+\mathcal E_T) / 2$ represent the average spectral width and mass-shift, $\Gamma_\Delta \equiv \Gamma_L-\Gamma_T$ and $\mathcal E_\Delta \equiv \mathcal E_L-\mathcal E_T$ represent the differences on the width and mass-shift between the longitudinal and transverse modes, and $\mathcal E^\prime \equiv \mathcal E-\mu_S$.

The following discussion is under the conjectured sequence of magnitudes that $\mathcal E_\Delta \sim \Gamma_\Delta \ll \Bar{\mathcal E} < \bar\Gamma \sim \mu_S \sim T \ll m < E_p$, which is justified, e.g., in the one-loop calculation on the $\phi$ meson spectral properties~\cite{Dong:2023cng}, where the $\mathcal E_\Delta$ and $\Gamma_\Delta $ turn out to be negligible compared to $\Bar{\mathcal E}$ and $\bar\Gamma$, and $\bar\Gamma$ is approximately half of the temperature. $T \ll m$ is apparent given the huge $K^\ast$ mass, and leads to $e^{E_p/T}\gg 1 \to n\ll 1$ with $n$ being the Bose distribution. The spectral function given by Eq. (\ref{Eq:Breit-Wigner}) is peaked around $\Omega\sim -\mu_S$. I therefore expand the Bose distribution $n(E_p+\Omega)$ around $\Omega\sim -\mu_S$ as $\partial n / \partial \varepsilon \sim [-1 + (\mu_S+\Omega)/T] n_\ast / T$ with $n_\ast \equiv n(E_p-\mu_S)$. Furthermore, according to Ref.~\cite{Dong:2023cng}, the mass-shift varies slowly with $|\mathbf p|$, which makes $\mathcal F \equiv 2 \partial_{\mathbf p^2} (E_p \mathcal E) \sim \mathcal O (\mathcal E/m) \ll 1$.

The transport coefficients in Eq.(\ref{Eq:CoefficientsEXACT}) can be estimated by first expanding the integrands till the second order of $\Omega$, conducting the integrals using Eq. (\ref{Eq:Moments}), further expanding the integrated results in terms of $\mathcal{O}(\Gamma_\Delta/ \bar \Gamma)$, $\mathcal O(T/m)$ or the other ratios at the similar order of magnitude according to the above conjectured sequence, and keeping only the leading contributions for each transport coefficient. The results are listed below as
\begin{widetext}
\begin{align}
    \label{Eq:CoefficientsEstimation}
    \vartheta^p_{sp.} \approx & - \frac{2 \mathfrak q m n_\ast}{E_p} \left[\frac{\mathcal E_\Delta}{\bar\Gamma}+\frac{(T-\Bar{\mathcal E})\Gamma_\Delta}{\bar\Gamma^2}+\frac{\bar\Gamma^2 + 4(\Bar{\mathcal E}-\mu_S)^2}{4 \bar\Gamma m} \frac{(\mu_S-T) \mathbf p^2}{m^3}\right],\nonumber\\
    \vartheta^p_{tt.} \approx & ~ 6 \mathfrak q n_\ast  \frac{m}{E_p} \frac{T-\Bar{\mathcal E}}{\bar\Gamma},\\
    \vartheta^u_{sp.} \approx & - 2 \mathfrak q n_\ast \left[\frac{\bar\Gamma^2 + 4(\Bar{\mathcal E}-\mu_S)^2}{4 \bar\Gamma m} \frac{(\mu_S-T) \mathbf p^2}{m^3} \frac{ \bar{\mathcal B} + \mu}{E_p}
    +\frac{T-\bar{\mathcal E}}{E_p} \left(\frac{\bar{\mathcal B} + \bar{\mathcal E}}{\bar \Gamma} \frac{\Gamma_\Delta}{\bar\Gamma} - \frac{\mathcal B_\Delta + \mathcal E_\Delta}{\bar\Gamma}\right)
    + \frac{\bar{\mathcal B}+\bar{\mathcal E}}{\bar \Gamma} \frac{\mathcal E_\Delta}{E_p} + \frac{\Gamma_\Delta}{4 E_p}
    \right],\nonumber\\
    \vartheta^u_{tt.} \approx & ~ 6 \mathfrak q n_\ast \left[\frac{T-\bar{\mathcal E}}{E_p} \left(\frac{\bar{\mathcal B}+\bar{\mathcal E}}{\bar \Gamma}\right) - \frac{\bar \Gamma}{4 E_p}\right], \nonumber\\
    \vartheta_\Delta \approx & -\frac{4 n_\ast E_p (T-\Bar{\mathcal E})}{\mathbf p^2} \frac{\mathcal E_\Delta}{\bar\Gamma},\nonumber
\end{align}    
\end{widetext}
where $\Bar{\mathcal B} \equiv 2 E_p \partial_{\mathbf p^2} (E_p \Bar{\mathcal E})$ and $\mathcal B_\Delta \equiv 2 E_p \partial_{\mathbf p^2} (E_p \mathcal E_\Delta)$.

One can see from Eq. (\ref{Eq:CoefficientsEstimation}) that $\mathcal O (\vartheta^u_{sp.}/\vartheta^p_{sp.}) \sim \mathcal O (\vartheta^u_{tt.}/\vartheta^p_{tt.}) \sim \mathcal O (T/m) \ll 1$, and $\mathcal O (\vartheta^p_{sp.}) \sim \mathcal O (\alpha_0 \vartheta^p_{tt.})$, $\mathcal O (\vartheta^u_{sp.}) \sim \mathcal O (\alpha_0 \vartheta^u_{tt.})$ since $\mathcal O (\alpha_0) \sim \mathcal O (\mathcal E_\Delta /T )$~\cite{Li:2022vmb}. $\vartheta_\Delta$ is sensitive to $\mathbf p^2$ and seems divergent at $\mathbf p^2 = 0$. Such a divergence is however an illusion. As demonstrated before, the longitudinal and transverse modes should degenerate for the zero momentum particles in the MRF, resulting a vanishing $\mathcal E_\Delta$ at $\mathbf p^2 = 0$ as well. On the other hand, according to equipartition theorem for the non-relativistic particles, $\overline{\mathbf p^2}\sim 3 T m$, which makes $\mathcal O (\vartheta_\Delta) \sim \mathcal O (\vartheta^p_{sp.})$ in the average sense. To conclude, under the quasi-particle approximation, the terms with the transport coefficient $\vartheta^p_{sp.}$, $\alpha_0\vartheta^p_{tt.}$, and $\vartheta_\Delta$ are the leading contributions to DITP. All these coefficients are, however, way smaller than the leading ones in the SITP effect~\cite{Li:2022vmb}.

Meanwhile, similar to the transport coefficients in the SITP effects~\cite{Li:2022vmb}, all the coefficients in Eq. (\ref{Eq:CoefficientsEstimation}) are T-odd, i.e., odd in the spectral width, whose physical meaning will be demonstrated in the next section.

\section{Summary}

In summary, I discover a new machinery, called diffusion induced tensor polarization (DITP),  contributing to the spin alignment of the strange (or charged) vector mesons with a magnitude proportional to the gradient of the strange chemical potentials, which is proven not small in the low energy HICs. The DITP effect is derived using the linear response theory, with the transport coefficients evaluated non-perturbatively, under certain assumptions, by employing the Ward-Takahashi identity. It turns out that these non-vanishing coefficients are mostly originating from the splitting between the longitudinal and transverse spectral properties, and are estimated to be much smaller than those of the SITP effect, under the quasi-particle approximation. The tensor structure of the DITP effect indicates that it contributes only to the \lq\lq local\rq\rq spin alignment in a static medium, and is hence most probably elusive in the current experiments where only the \lq\lq global\rq\rq spin alignments are measured. However, this should not undermine the significance of the DITP effect. Since the strange chemical potential is proven positively dependent on the baryon chemical potential for maintaining strange neutrality, DITP should also increase with the gradient of the baryon chemical potential, while the latter shall be drastically enhanced during the first order phase transition. Once the \lq\lq local\rq\rq spin alignments are measured, the DITP effect might be a novel probe of the QCD first order phase transition.

Future efforts will be made from two aspects. Theoretically, the transport coefficients defined in Eq. (\ref{Eq:CoefficientsEXACT}) will be calculated with the spectral properties being evaluated using the FRG formalism. Phenomenologically, I shall generate the temperature and chemical potential distribution on the freeze-out hyper-surface from realistic simulations based on either the hydrodynamic or the transport models.

This study bridges two key and interesting problems in the high energy nuclear physics, i.e., the spin alignment of the vector mesons and the properties of the QCD phase transition, and hence opens new perspectives for improving our knowledge on the both sides.

\appendix 
\section{Properties of Projectors}
\label{sec:AppdxProjector}
The longitudinal and transverse projectors defined as
\begin{equation}
    \label{Eq:LTProjector}
    P_L^{\mu\nu} \equiv -\frac{u_\perp^\mu u_\perp^\nu}{u_\perp^2}, ~~~~
    P_T^{\mu\nu} \equiv P^{\mu\nu} - P_L^{\mu\nu},
\end{equation}
with $u_\perp \equiv P\cdot u$ for $u$ representing the flow velocity, take the components not only perpendicular to the particle momentum, but also parallel and perpendicular to the flow velocity, respectively, and satisfy the following identities
\begin{align}
    \label{Eq:ProjectorIdentities}
    P\cdot P_L &= P_L \cdot P_L = -P_L,~~P\cdot P_T = P_T \cdot P_T = -P_T,\nonumber\\
    P_L\cdot P_T &=0,~~ P^\mu_{L\mu}=-1, ~~P^\mu_{T\mu}=-2, ~~P^\mu_{\mu}=-3, \nonumber\\
    & p\cdot P(p)=p\cdot P_L(p) = p\cdot P_T(p)=0, \nonumber\\
    & P\cdot u = P_L \cdot u = u_\perp, ~~ u\cdot P_T = 0, \\
    & P_L^{\alpha\beta} u_\perp^\gamma = P_L^{\beta\gamma} u_\perp^\alpha = P_L^{\gamma\alpha}u_\perp^\beta =\cdots, \nonumber\\
    & P^{\alpha\rho}_a P^{\beta\sigma}_b \partial_{p^\mu} P_{\rho\sigma}=0~~{\rm for~} a,b=L~{\rm or}~T, \nonumber\\
    P^{\alpha\rho}_a P^{\beta\sigma}_b & \partial_{p^\mu} P^L_{\rho\sigma}= -P^{\alpha\rho}_a P^{\beta\sigma}_b \partial_{p^\mu} P^T_{\rho\sigma}~~{\rm for~} a,b=L~{\rm or}~T.\nonumber
\end{align}

The physical meanings of $P_T$ is more transparent in the medium rest frame (MRF), where $u=(1,\mathbf 0)$ and 
\begin{equation}
    \label{Eq:ProjectorInMRF}
    P^{\rm MRF}_L = \frac{p_0^2}{p^2} \left(
    \begin{tabular}{c c}
      $\mathbf v^2$  & $\mathbf v$ \\
      $\mathbf v$  &  $\hat{\mathbf p} \hat{\mathbf p}$
    \end{tabular}\right),~
    P^{\rm MRF}_T = \left(
    \begin{tabular}{c c}
      $0$  & $0$ \\
      $0$  &  $\mathbf I -\hat{\mathbf p} \hat{\mathbf p}$
    \end{tabular}\right),
\end{equation}
with $\hat{\mathbf p}\equiv \mathbf p / |\mathbf p|$ being the unit 3-vector in the direction of $\mathbf p$, and $\mathbf v \equiv \mathbf p /p^0$ being the {\it off-shell} particle 3-velocity. In the MRF, $P^{\rm MRF}_T$ takes the spatial components perpendicular to $\mathbf p$, this is the reason why $P_T$ is regarded \lq\lq transverse\rq\rq. 

Eq. (\ref{Eq:ProjectorInMRF}) exhibits a key property of $P_T$, i.e., $P_T^{\rm MRF}$ does not depend on $p^0$, neither depend on the particle mass. Furthermore, since a general $P_T$ is linked with $P_T^{\rm MRF}$ by a Lorentz boost with the frame velocity equal to $\mathbf u$, i.e., $P_T = \Lambda (\mathbf u)\cdot P_T^{\rm MRF} \cdot \Lambda^T (\mathbf u)$, it does NOT depend on the particle mass in an ARBITRARY frame either. Hence, 
\begin{equation}
    \label{Eq:PTOffShell}
    P_T(p)=P_T(\tilde p)\equiv \tilde P_T,   
\end{equation}
with $\tilde p \equiv (E_p, \mathbf p)$ is the on-shell 4-momentum. The connection between $P_L(p)$ and $\tilde P_L$ is not as simple as Eq. (\ref{Eq:PTOffShell}). However, if sandwiched by the on-shell polarizers or projectors, $P_L (p)$ and $\tilde P_L$ obey the simple relations as follows:
\begin{eqnarray}
    \label{Eq:PLOffShell}
    \epsilon_0^{\langle \alpha} P^L_{\alpha\beta}(p) \epsilon_0^{\beta\rangle} &=& \epsilon_0^{\langle \alpha} \tilde P^L_{\alpha\beta} \epsilon_0^{\beta\rangle}
    \left(1+ \frac{\Omega^2 \kappa^2}{m^2 p^2}\right), \nonumber\\
    \tilde P^{\alpha\beta} P^L_{\alpha\beta}(p)  &=& \tilde P^{\alpha\beta} \tilde P^L_{\alpha\beta}
    \left(1+ \frac{\Omega^2 \kappa^2}{m^2 p^2}\right),
\end{eqnarray}
with $\Omega \equiv u\cdot (p - \tilde p)$, $\kappa^2 \equiv (u\cdot \tilde p)^2-m^2 =  (u\cdot p)^2-p^2$. 
Similarly,
\begin{eqnarray}
    \label{Eq:UperpOffShell}
    \epsilon_0 \cdot u_\perp (p) &=& -\epsilon_0 \cdot u \left(1-\frac{p\cdot u \Omega}{p^2}\right), \\
    \tilde P \cdot u_\perp (p) &=& -\tilde P \cdot u \left(1-\frac{p\cdot u \Omega}{p^2}\right). \nonumber
\end{eqnarray}
Futhermore, the on-shell projectors obey the identity:
\begin{equation}
    \label{Eq:PTPLOnshell}
    \epsilon_0^{\langle \alpha} \tilde P^L_{\alpha\beta} \epsilon_0^{\beta\rangle} = -\epsilon_0^{\langle \alpha} \tilde P^T_{\alpha\beta} \epsilon_0^{\beta\rangle}.
\end{equation}

In the end, let us look into the momentum derivative on $P_T$ presenting in Eq. (\ref{Eq:ThetaFactor}). Given Eq. (\ref{Eq:ProjectorInMRF}), in the MRF, 
\begin{align}
    \label{Eq:DPTInMRF}
    & P^{\alpha\rho}_a P^{\beta\sigma}_b \partial_{p_\mu} P^T_{\rho\sigma} \nonumber\\
    = & - P^{\alpha i}_a P^{\beta j}_b \left( \frac{\delta^\mu_i p_j}{\mathbf p^2} + \frac{\delta^\mu_j p_i}{\mathbf p^2} + 2 \frac{p_i p_j p^\mu}{\mathbf p^4}\right)\delta_{\mu \neq 0} \\
    = & p^0 \left( \frac{P^{\alpha \mu}_a P^{\beta 0}_b}{\mathbf p^2} + \frac{P^{\alpha 0}_a P^{\beta \mu}_b }{\mathbf p^2} - 2 \frac{P^{\alpha 0}_a P^{\beta 0}_b p^0 p^\mu}{\mathbf p^4}\right)\delta_{\mu \neq 0}. \nonumber
\end{align}
Eq. (\ref{Eq:DPTInMRF}) can be extended to an arbitrary frame by conducting the following transformation: $p^0 \to u\cdot p$, $\mathbf p^2 \to \kappa^2$, $P^{\mu 0}_a \to P^{\mu\nu}_a u_\nu = 0$ (for $a=T$) or $u_\perp^\mu$ (for $a=L$), and $V^\mu \delta_{\mu\neq 0} \to \Delta^{\mu\nu} V_\nu$ with $V$ representing an arbitrary 4-vector and $\Delta ^{\mu\nu} \equiv g^{\mu\nu}-u^\mu u^\nu$ being the flow projector that takes the components perpendicular to $u$. Under these transformations, Eq. (\ref{Eq:DPTInMRF}) are thus extended to
\begin{align}
    \label{Eq:DPT}
    P^{\alpha\sigma}_T P^{\beta\rho}_T \partial_{p_\mu} P^T_{\rho\sigma} &= 0, \nonumber\\
    P^{\alpha\sigma}_L P^{\beta\rho}_T \partial_{p_\mu} P^T_{\rho\sigma} &= u\cdot p \left( \frac{u_\perp^\alpha \tilde P^{\beta \nu}_T \Delta^\mu_\nu}{\kappa^2}\right)= u\cdot p \left( \frac{u_\perp^\alpha \tilde P^{\beta \mu}_T}{\kappa^2}\right), \nonumber\\
    P^{\alpha\sigma}_T P^{\beta\rho}_L \partial_{p_\mu} P^T_{\rho\sigma} &= u\cdot p \left( \frac{u_\perp^\beta \tilde P^{\alpha \mu}_T}{\kappa^2}\right),\\
    P^{\alpha\sigma}_L P^{\beta\rho}_L \partial_{p_\mu} P^T_{\rho\sigma} &= 2u\cdot p \left(\frac{u_\perp^{\nu}}{\kappa^2}  +  \frac{u_\perp^2 u\cdot p p^\nu}{\kappa^4}\right)P_L^{\alpha\beta}\Delta^\mu_\nu = 0. \nonumber
\end{align}

\acknowledgments
The author thanks Shuai Y.F. Liu and Min He for inspiring and helpful discussions, especially for those on the analyticity of the spectral functions with Shuai Y.F. Liu. This research is supported by NSFC No. 12105129 and NSFC No. 12205090.
\bibliography{ref}

\end{document}